\pdfoutput=1

\documentclass{iau}

\usepackage{graphicx}
\usepackage{multicol}
\usepackage{natbib}
\usepackage{hyperref}
\hypersetup{
  bookmarks=false,         
  bookmarksopen=false,     
  pdffitwindow=true,       
  pdftitle={AGN torus properties with WISE},    
  pdfsubject={Proceedings IAU Symposium No. 304, 2013},    
  pdfauthor={Nikutta et al.},   
  colorlinks=true,         
  linkcolor=blue,          
  citecolor=blue,          
  urlcolor=blue,           
  linktocpage=true
}

\newcommand\C{\textsc{Clumpy}}
\newcommand\tI{\hbox{type~1}}
\newcommand\tII{\hbox{type~2}}
\def\c[#1,#2]{\hbox{${\rm W}#1-{\rm W}#2$}}  

\title[AGN torus properties with WISE] 
{AGN torus properties with WISE}

\author[R. Nikutta et al.]   
{Robert Nikutta$^{1,2}$, Maia Nenkova$^3$, \v{Z}eljko~Ivezi\'{c}$^{4}$, Nicholas~Hunt-Walker$^{4}$ \and Moshe Elitzur$^{2}$
}

\affiliation{$^1$Departamento de Ciencias F\'isicas, Universidad Andr\'{e}s Bello, \\
Av. Rep\'{u}blica 252, Santiago, Chile \\ email: {\tt robert.nikutta@gmail.com} \\[\affilskip]
$^2$Department of Physics \& Astronomy, University of Kentucky, \\
Lexington, KY 40506, USA \\ email: {\tt moshe@pa.uky.edu} \\[\affilskip]
$^3$School of English and Liberal Studies, Seneca College, \\ 
Toronto, ON, M2J 2X5, Canada \\ email: {\tt Maia.Nenkova@senecacollege.ca} \\[\affilskip]
$^4$Astronomy Department, University of Washington, \\ 
Box 351580, Seattle, WA 98195-1580, USA \\ email: {\tt ivezic@astro.washington.edu, nhuntwalker@gmail.com}}

\pubyear{2013}
\volume{304}  
\pagerange{1--4}
\jname{Multiwavelength AGN Surveys and Studies}
\editors{A. Mickaelian, F. Aharonian \& D. Sanders, eds.}
\begin{document}

\maketitle

\begin{abstract}
  The Wide-field Infrared Survey Explorer (WISE) has scanned the
  entire sky with unprecedented sensitivity in four infrared bands, at
  3.4, 4.6, 12, and 22$\mu{\rm m}$. The WISE Point Source Catalog
  contains more than 560 million objects, among them hundreds of
  thousands of galaxies with Active Nuclei (AGN). While \tI\ AGN,
  owing to their bright and unobscured nature, are easy to detect and
  constitute a rather complete and unbiased sample, their \tII\
  counterparts, postulated by AGN unification, are not as
  straightforward to identify. Matching the WISE catalog with known
  QSOs in the Sloan Digital Sky Survey we confirm previous
  identification of the \tI\ locus in the WISE color space. Using a
  very large database of the popular \C\ torus models, we find the
  colors of the putative \tII\ counterparts, and also, for the first
  time, predict their number vs. flux relation that can be expected to
  be observed in any given WISE color range. This will allow us to put
  statistically very significant constraints on the torus
  parameters. Our results are a successful test of the AGN unification
  scheme.
\end{abstract}

\firstsection 
\section{Introduction}

\noindent AGN unification postulates that the average make-up of \tI\
and \tII\ AGN populations is intrinsically identical. The central
engine of AGN is a hot accretion disk, with gas spiraling down towards
a supermassive black hole. Some distance away it is surrounded by a
dusty structure, geometrically and optically thick, and concentrated
toward the equatorial plane of the system. This axisymmetric ``torus''
provides open lines of sight towards the nucleus along polar
directions, but simultaneously blocks the view in edge-on
directions. It gives rise to the \tI/\tII\ dichotomy found in
observations: there are AGN with both broad and narrow emission lines
(\tI) in their spectra, and those with only narrow lines (\tII). In
the latter, the view of the gas clouds in the Broad Line Region is
simply blocked by the torus. Of course, broad line photons emitted in
axial directions and then scattered off-axis led to their discovery in
the polarized light from type-2 AGN \citep{AntonucciMiller1985},
ultimately inspiring the unification paradigm
\citep{UrryPadovani1995}.

\section{CLUMPY torus models}

\noindent \citet{KrolikBegelman1988} discussed very early severe
dynamical problems in sustaining the torus scale height, derived from
\tI/\tII\ number statistics, if the dust were smoothly
distributed. The problem could be remedied if the dust were contained
in optically thick clouds instead, shielding itself from the energetic
AGN radiation. Most early models of AGN torus emission employed
smooth-density tori \citep[e.g.,][]{GranatoDanese1994,
  EfstathiouRowan-Robinson1995, Gratadour+2003}. With few exceptions
\citep[e.g.,][]{Fritz+2006} they predicted spectral features
incompatible with later observations, among them the lack of deep
silicate absorption features \citep[e.g.,][]{Hao+2007,Wu+2009}, and
the discovery of silicate emission in \tII s
\citep{Mason+2009,Nikutta+2009}.

Today, VLTI observations in the mid-IR and X-ray variability studies
strongly support the clumpy regime \citep[e.g.,][]{Jaffe+2004,
  Poncelet+2006, Tristram+2007, Raban+2009, Risaliti+2002,
  Rivers+2011, MKN2013, NKM2014}. The development of dust radiative
transfer in clumpy media \citep{Nenkova+2002, Nenkova+2008a,
  Nenkova+2008b, Hoenig+2006, Schartmann+2008, Stalevski+2012}
reconciled theory and observations, although some problems remain
\citep{ElitzurShloman2006}. Our \C\ formalism
\citep{Nenkova+2002,Nenkova+2008a,Nenkova+2008b} was the first
self-consistent clumpy torus model. Exploiting IR scaling relations
\citep{IE97}, we are able to provide a large set of comprehensive
model SEDs for the
community\footnote{\url{http://www.pa.uky.edu/clumpy/}}; they are
being used extensively in SED fitting
\citep[e.g.,][]{AsensioRamos2009,Alonso-Herrero+2011,
  Deo+2011,Malmrose+2011,Mason+2013}.

\section{AGN in the WISE catalog}

\noindent \C\ SEDs have been computed for a very large parameter
volume. Each possible combination of parameter values yields a
different SED. In the observational domain, however, samples of AGN
detected in the IR were rather small. It is only with the arrival of
WISE \citep{Wright+2010}, and the release of its vast point source
catalog\footnote{\url{http://wise2.ipac.caltech.edu/docs/release/allsky/}},
that the IR colors of many more AGN are available. Despite WISE's
rather low angular resolution it can be expected that the NIR/MIR
fluxes of AGN are dominated by the nuclear dust emission, especially
in type-1 QSOs. WISE scanned the entire sky with great sensitivity in
four bands. A photometric color between two bands, say 1 and 2, is
simply the magnitude difference recorded between these two bands,
i.e. \c[1,2]. Since the actual measurements are fluxes, the color is
also \mbox{$\c[1,2] = 2.5 \log (f_2/f_1)$}, where for the WISE catalog
the $f_i$ are the fluxes calibrated in the Vega photometric system.

\begin{figure}[t]
\begin{center}
 \includegraphics[width=0.8\hsize]{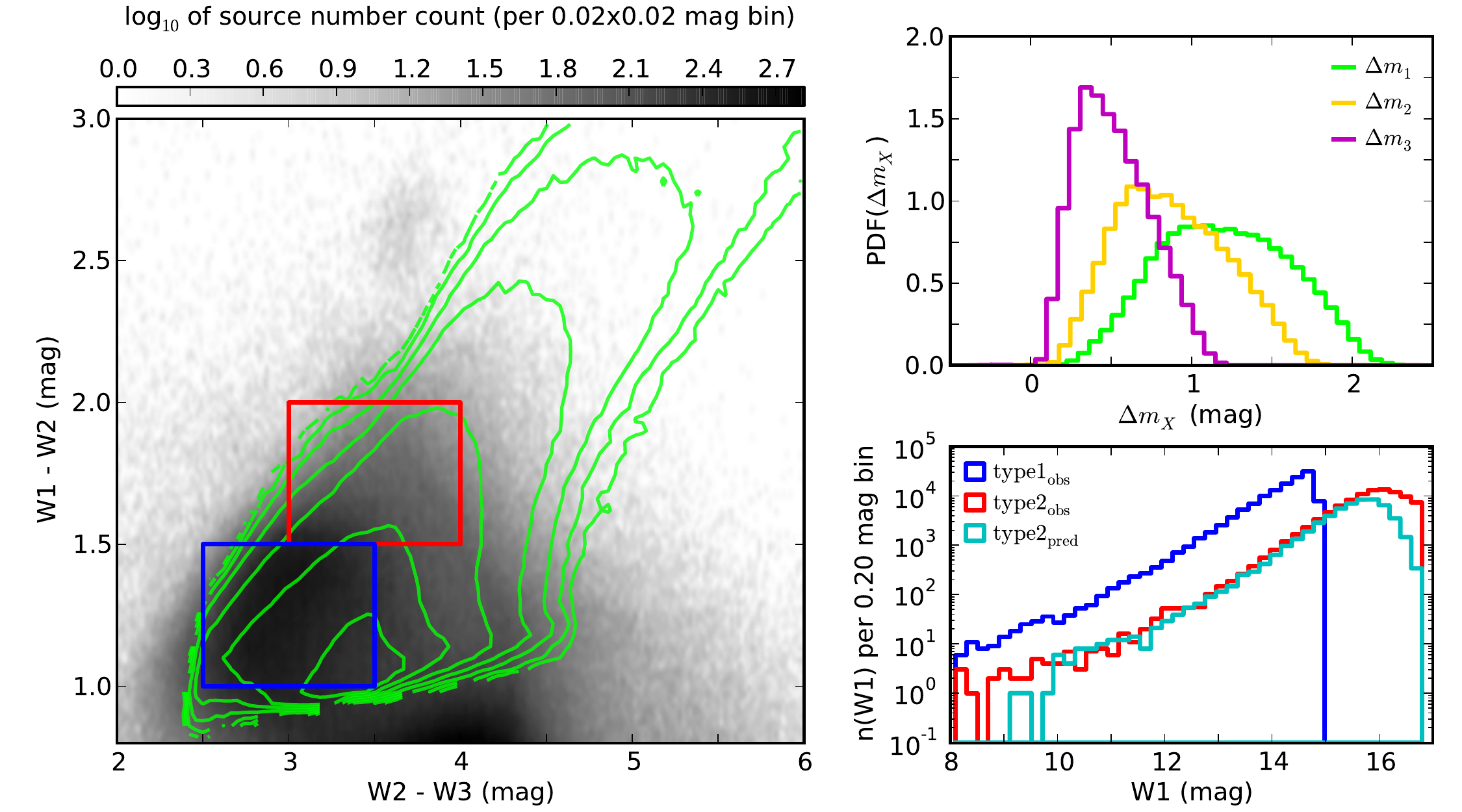} 
 \caption{Testing unification with WISE observations and \C\
   models. The figure is from the forthcoming \citet{WISE2}. {\em
     Left:} Red WISE sources in a CC diagram, shown with a gray
   scale. All sources satisfy $5\sigma$ detection limits. The blue box
   outlines the locus of QSOs \citep{Yan+2013}; their number is
   $\approx$108,000. The population of putative \tII\ counterparts is
   distributed throughout the CC diagram -- one possible locus is
   highlighted with a red box. Green contour lines show the number
   density of all \C\ models that offer some viewings with colors in
   the blue box, i.e. \tI-compatible colors (but all viewings to such
   models contribute); this is the ``model sample''. {\em Upper
     right:} Mean (per model) dimming $\Delta m_i$ of \C\ model
   fluxes, when switching from \tI\ compatible viewings (colors in the
   blue box) to viewings in the red box. The green, orange, magenta
   histograms show the distribution of model dimmings in bands 1, 2,
   3, respectively. {\em Lower right:} Distribution of W1 magnitudes
   of the sources. The blue histogram is of all objects in the blue
   \tI\ box from the left panel. The red histogram is of the sources
   in the red box. The cyan histogram is of all \tI\ sources after
   dimming them using knowledge from \C\ models (see lower right
   panel).}
   \label{fig1}
\end{center}
\end{figure}
The left panel of Figure \ref{fig1}, from the forthcoming
\citet{WISE2}, shows the distribution of all WISE sources with
significant IR excess in the \c[1,2] vs. \c[2,3] color-color (CC)
diagram. The logarithmic gray scale corresponds to the number of
sources per $0.02\times0.02$~mag color bin. All $\approx\!\!
1.3\!\cdot\!10^6$ sources satisfy $5\sigma$ detection limits. Except
for the smallest \c[1,2] colors at the bottom of the panel (local
IR-luminous sources reside there, such as Young Stellar Objects), this
CC diagram is dominated by extragalactic sources. \citet{Yan+2013}
have previously identified the locus of \tI\ QSOs in this color-color
space. Cross-checking with QSOs in the Sloan Digital Sky Survey
(SDSS), we confirm this locus, and outline it roughly with the blue
rectangle in Fig.~\ref{fig1}. This box is dominated by unobscured \tI\
QSOs, and contains $\sim\! 108,000$ sources. The distribution of their
W1 magnitudes is shown as a blue histogram in the lower right
panel. The cut-off at faint magnitudes is caused by the WISE detection
sensitivity.

\section{Testing AGN unification}

\noindent A given \tI\ AGN, if it could be rotated to a \tII\
perspective, would not only change its optical type, but also its IR
appearance. Observed with WISE, it would register dimmer in all bands,
and redder in all colors. This putative \tII\ counterpart would be
more difficult to detect, as it might become too faint in one of
WISE's bands and simply drop out from the sample. We can expect that
the sub-sample of \tII\ counterparts, drawn from the same overall AGN
population as \tI s, will distribute in regions of the CC diagram that
are redder than the QSO locus. Where exactly though, we can not know
purely from observations, because each individual \tII\ counterpart
will have different internal properties, leading to varying amounts of
dimming in each band, and thus to different color shifts. Although we
can expect a typical color shift when switching from \tI\ to \tII\
viewing, we have no observational handle on the dimming per band that
caused such color shifts.

One possible location of \tII\ counterparts is marked with a red box
in Fig. \ref{fig1}. The red histogram n(W1) shows their magnitude
distribution in the lower right panel. For all putative counterparts
that fall within the red box, we know the average color shift, but not
the magnitude dimming that is required per WISE band. This missing
knowledge can be provided by the \C\ models. The green contour lines
in the left panel show the number density of all \C\ models that offer
some viewings with colors in the blue box, i.e. \tI-compatible
colors. Note that all viewings to such models contribute to the
density contours (i.e. \tI s and their \tII\ counterpart
viewings!). This is the ``model sample''.

To estimate the amount of dimming per WISE band required for an
overall color shift from the blue to a red box, we pick from the model
sample those models that also have some viewings with colors in the
red box. For each selected model we then average their \tI-viewing
magnitudes and their \tII-viewing magnitudes separately, and the
difference between these averages is the dimming (determined
separately per WISE band). We repeat this procedure for every
qualifying model. The distributions of dimmings obtained in this
fashion are shown in the upper right panel of Fig. \ref{fig1}, for
WISE bands 1, 2, 3 (in green, orange, and magenta, respectively).

This final step in testing unification is to apply the model-based
dimming in each band to every observed QSO source. To this end we
randomly draw {\em correlated} dimmings from the distributions of
dimming magnitudes, for every source from the blue box. After dimming
the source by the obtained dimming magnitudes we determine if it
survives the WISE $5\sigma$ faint limits (per band), and discard it
otherwise. The distribution of surviving dimmed \tI\ sources is shown
as a cyan histogram in the lower right panel. The agreement with the
red histogram (observed putative \tII\ sources) is striking. Note that
the cyan histogram is the {\em predicted} absolute number of \tII\ AGN
to be observed with a survey of given sensitivity.

Repeating this algorithm on all other potential \tII\ regions in the
color space (red boxes) will identify the area in the color space that
harbors the population of \tII\ counterparts. Then, knowing this
area's extent in color space we will be able to constrain the
permitted values of AGN torus parameters -- based on a $10^6$ sample
of objects!  We will report all results in greater detail in
\citet{WISE1,WISE2}.

\begin{multicols}{2}

\end{multicols}

\end{document}